\documentclass[aps, pra, twocolumn, superscriptaddress, amsmath, 
tightenlines, longbibliography]{revtex4-2}

\usepackage{amssymb}
\usepackage{amsmath}
\usepackage{dcolumn}
\usepackage{graphicx}
\usepackage{mathrsfs}
\usepackage{appendix}
\usepackage{graphicx}
\usepackage{booktabs}
\usepackage{units}

\def \hide#1{}

\usepackage{url}
\usepackage[colorlinks]{hyperref}
\hypersetup{%
	plainpages=true,
	breaklinks=true,       
	hypertexnames=false,  
	pageanchor=true,
	colorlinks=true,
	linkcolor={blue},
	citecolor={red},
	urlcolor={blue},
	anchorcolor={black}
}

\begin{document}

	\title{Chiral doublon bound states in a synthetic topological waveguide}

	\author{Ying Xia}
	\affiliation{Shaanxi Province Key Laboratory of Quantum Information and Quantum Optoelectronic Devices,
	School of Physics, Xi’an Jiaotong University, Xi'an 710049, People’s Republic of China}
	
	\author{Xin Wang}\email{wangxin.phy@xjtu.edu.cn}
	\affiliation{Shaanxi Province Key Laboratory of Quantum Information and Quantum Optoelectronic Devices,
	School of Physics, Xi’an Jiaotong University, Xi'an 710049, People’s Republic of China}

\date{\today}

\begin{abstract}
Nonlinear waveguide QED serves as a platform for studying correlated photon dynamics, where its interplay with topological bound states can give rise to unconventional phenomena. We investigate a triangular ladder waveguide in which a pair of emitters forms a doublon bound state through strong nearest-neighbor interactions. By mapping the doublon dynamics onto an effective SSH chain, we show that the system is equivalent to an effective emitter coupled to a topological doublon bath. Using the vacancy-like dressed state approach, we demonstrate the existence of a chiral bound state with a node at the coupling site and an exponentially localized wave function inherited from the SSH edge states. Extending this picture to two emitter pairs, we derive an effective four-body interaction whose strength is determined by the relative chirality and the separation parity of the bound states. When the two bound states face each other, coherent Rabi oscillations emerge; when they are arranged back-to-back or have an even separation, the interaction vanishes. Our results provide a route toward engineering tunable many-body interactions between correlated photon pairs and realizing nonlinear quantum networks based on flying doublons.
\end{abstract}

\maketitle

\section{introduction}
Waveguide quantum electrodynamics (QED) provides a versatile platform for exploring the interplay between quantum emitters and structured photonic environments, in which the photon-mediated interaction can be engineered through the dispersion and dimensionality of photonic reservoirs~\cite{liaophoton2016,Lodahl2015,Bliokhtransverse2015,ChangDE2018,Stewart2020,Trainiti2016}. In conventional linear waveguides, the dynamics is mainly restricted to the single-excitation regime~\cite{Karplus1951,LiaoJieQiao2010,ZhouLan2008}, leading to a variety of intriguing phenomena, including non-Markovian emitter dynamics~\cite{Ramos2016,GonzalezTudela2017,GonzlezTudela2018}, chiral photon transport~\cite{DeBernardis2023,Wangchiral2022,Tang2022,Dulei2022,lodahlchiral2017}, and emitter-photon bound states induced by modified photonic densities of states~\cite{GarcaElcano2020,ShiTao2016,LiuYanbing2016}. These bound states, formed when an excitation becomes partially localized around an emitter owing to band gaps or singularities in the photonic spectrum~\cite{John1990,Kofman1994,Lambropoulos2000}, provide a powerful mechanism for controlling light-matter interactions beyond the Markovian approximation~\cite{Calajo2016,JohnQuang1994,Sundaresan2019}. However, the absence of intrinsic photon-photon interactions in linear photonic systems limits the possibility of realizing correlated many-photon phenomena and engineering collective quantum states.

Introducing nonlinear interactions into photonic systems provides a promising route to extending waveguide QED beyond the single-excitation manifold~\cite{2LiaoJieQiao2010,Shen2007,Chang2014,Dorfman2016,Sheremet2023,Poshakinskiy2021,Roy2017,Mahmoodian2018}. Strong photon-photon interactions can generate correlated few-photon states, including tightly bound photon pairs—doublons~\cite{MarquesY2021,Winklerrepulsively2006,Piil2007,WangYM2010,DiLibertoM2016}. Unlike individual photons, doublons behave as emergent quasiparticles whose dynamics are governed by effective band structures of the underlying nonlinear lattice~\cite{Salerno2020,Gorlach2017,Liang2018,Stepanenko2020}. In nonlinear waveguides, doublons give rise to several phenomena with no counterpart in linear systems: supercorrelated radiance~\cite{Wangzhihai2020,Lijiaqi2025}, doublon bound states in the continuum~\cite{Rieck2025,Zhangxiaojun2026}, and long-range interactions between distant emitter pairs~\cite{Wangxinlongrange,Wangnonlinear2026}. Recently, the interplay between doublon physics and topological band structures has attracted increasing interest~\cite{Berti2022,Flannigan2020,Lyubarov2019,Stepanenko2022}.

Topological photonics offers another route for engineering light-matter interactions in waveguide QED~\cite{Barik2018,Bettles2017,Perczel2017,Ozawa2019,gaoquantum2024}. Coupling quantum emitters to topological photonic reservoirs gives rise to unconventional bound states~\cite{Vega2021,Garmon2026}, topologically protected quantum entanglement~\cite{Luo2024,Rechtsman2016,Mittal2016,Wangmchelle2018}, and chiral light-matter interfaces~\cite{Cheng2022,zhuhai2025}. The Su-Schrieffer-Heeger (SSH) chain~\cite{SuWP1979}, with its symmetry-protected midgap edge states, provides the simplest platform for realizing these phenomena~\cite{Sirker2014,Bello2019}. In particular, vacancy-like dressed states (VDSs) have been identified in emitter-SSH systems, in which the emitter effectively acts as a vacancy that hosts a topologically protected chiral bound state~\cite{Leonforte2021}. Nevertheless, the study of doublon-induced chiral bound states in topological waveguide QED remains limited.

In this work, we show that a nonlinear triangular ladder serves as a synthetic topological waveguide for doublon excitations. By exploiting strong nearest-neighbor interactions, we map the two-photon dynamics onto an effective SSH lattice, where the doublon bands exhibit nontrivial topology and host midgap states. Coupling quantum emitters to this synthetic doublon bath gives rise to vacancy-like dressed states whose wave functions inherit the chiral sublattice selectivity of the underlying SSH edge modes. We further show that these chiral doublon bound states mediate controllable dipole-dipole interactions between distant emitter pairs, yielding an effective four-body interaction governed by topology and chirality. Our results provide an approach for designing correlated-photon-mediated many-body quantum interactions and offer a new perspective on nonlinear topological waveguide QED.
\section{Model}
\label{sec:Model}
\begin{figure*}[ht]
	\centering \includegraphics[width=\textwidth]{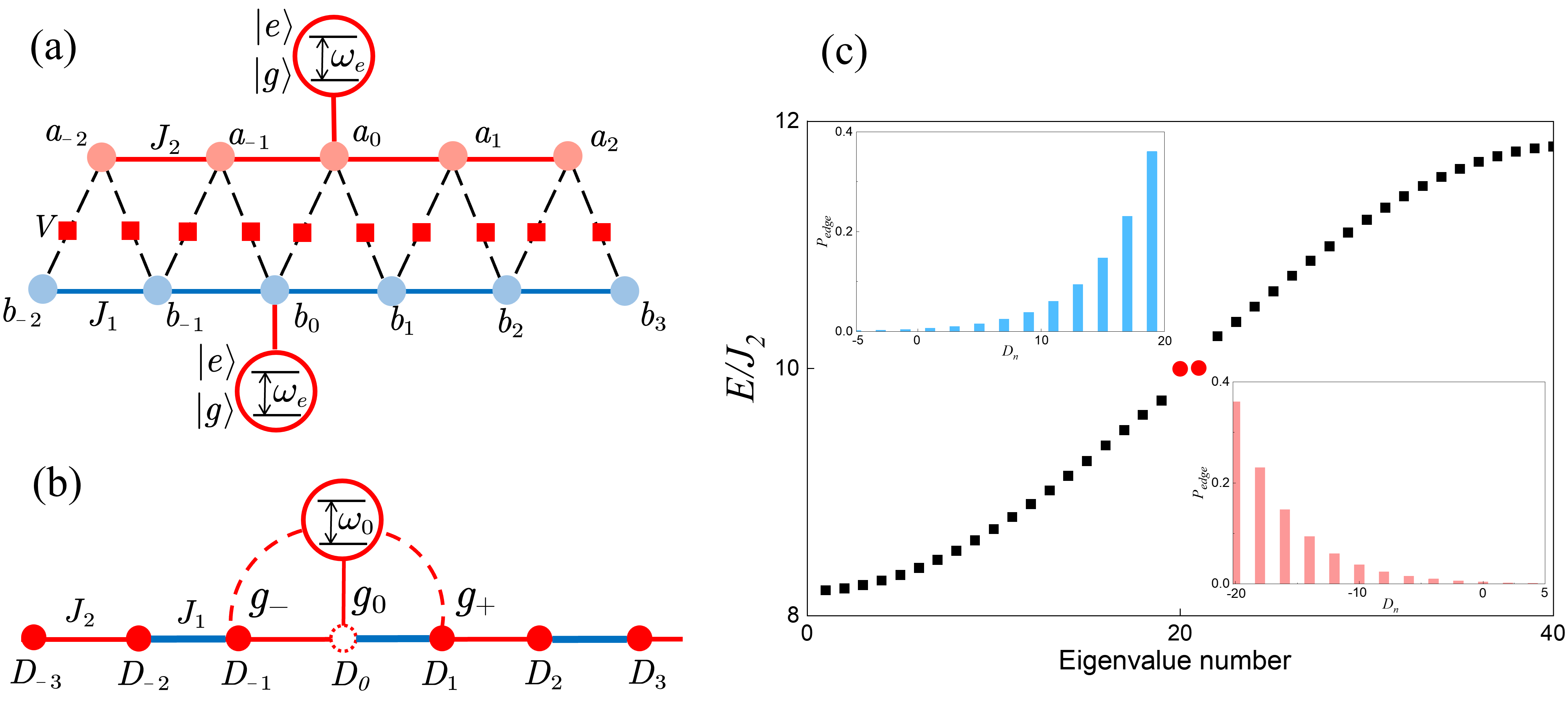}
	\caption{(a) A pair of emitters coupled to the triangular ladder model. $J_1$ and $J_2$ denote the nearest-neighbor hopping amplitudes along the upper and lower legs, respectively, and $V$ is the nearest-neighbor interaction on the rungs. 
	(b) Effective dual SSH chain with an emitter coupled to the central site. 
	(c) Two-body energy spectrum of the triangular ladder model. The insets show the two edge states of the effective dual doublon model.
	The parameters are $V=10J_2$, $J_2=1$, $J_1=0.8J_2$, and $L=40$ sites.
	}

	\label{fig1}
\end{figure*}

As illustrated in Fig.~\ref{fig1}(a), we consider a pair of quantum
emitters, each with transition frequency $\omega_e$, coupled to a
triangular ladder waveguide. We begin by briefly reviewing the description of tightly bound two-particle states on a lattice with nearest-neighbor interactions. In the two-excitation subspace, the bath Hamiltonian is given by ($\hbar = 1$)
\begin{align}
	H_B
	&=
	\sum_n
	\Big[
	-J_2 a_n^\dagger a_{n+1}
	-J_1 b_n^\dagger b_{n+1}
	+\mathrm{H.c.}
	\nonumber\\
	&\hspace{2.5em}
	+V\left(
	a_n^\dagger a_n b_n^\dagger b_n
	+
	a_n^\dagger a_n b_{n+1}^\dagger b_{n+1}
	\right)
	\Big],
\end{align}
where $J_i$ ($i=1,2$) denotes the hopping amplitude along each leg of the ladder, $a_n$ ($b_n$) is the photon annihilation operator
on sublattice $a$ ($b$) in the $n$th unit cell, and $V$ describes the
nearest-neighbor photon interaction on the corresponding lattice
bonds.

In the strong-interaction regime $V\gg J_1,J_2$, the low-energy dynamics
is dominated by tightly bound two-particle states of the form
$|D_n\rangle=a_{n_1}^\dagger b_{n_2}^\dagger|0\rangle$, where $n_1$ and
$n_2$ label the two sites connected by an interaction bond. We refer to
these two-body bound states as doublons~\cite{MarquesY2021,
Winklerrepulsively2006,Piil2007,WangYM2010,DiLibertoM2016,Salerno2020,
Gorlach2017,Liang2018,Stepanenko2020}. The center of mass of a doublon is located at the midpoint of a lattice bond, which defines an effective dual lattice on which the doublon dynamics takes place~\cite{Salerno2020}. 

With different hopping amplitudes $J_1$ and $J_2$ along the two legs,
the doublon dynamics on the effective dual lattice acquires alternating
hopping amplitudes and can be mapped to a Su-Schrieffer-Heeger (SSH)
model~\cite{SuWP1979,Sirker2014}, as shown in Fig.~\ref{fig1}(b). The resulting two-body spectrum consists of two doublon bands separated by a gap that hosts two midgap edge states, as shown in Fig.~\ref{fig1}(c). In the topological regime $J_1<J_2$, the corresponding edge-state wave functions, shown in the insets of Fig.~\ref{fig1}(c), are localized at opposite boundaries and decay exponentially into the bulk.

We next consider a pair of quantum emitters locally coupled to the two legs of the triangular ladder at the same unit cell $n_0$. The total Hamiltonian is given by
\begin{equation}
	H=H_B+\frac{\omega_e}{2}\left(\sigma_1^z+\sigma_2^z\right)
	+g\left(a_{n_0}\sigma_1^\dagger+b_{n_0}\sigma_2^\dagger+\mathrm{H.c.}\right),\label{originalH}
\end{equation}
where $g$ is the coupling strength, $\sigma_i^z$ and $\sigma_i^\dagger$ ($\sigma_i$) are the Pauli operators of the emitter $i$. We tune the emitter transition frequency to half of the edge-state energy, i.e., $\omega_e=E_{\mathrm{edge}}/2$. In the two-excitation subspace and under the weak-coupling condition $g \ll J_i,V$, the wave function of the total system can be expanded as
\begin{align}
	|\psi(t)\rangle &= \Bigg\{
	c_e(t)\sigma_1^\dagger\sigma_2^\dagger
	+ \sum_{n_1,n_2} c_D(n_1,n_2;t)
	a_{n_1}^\dagger b_{n_2}^\dagger
	\nonumber \\
	&+\sum_k \left[
	c_{1k}(t)\sigma_1^\dagger b_k^\dagger
	+c_{2k}(t)\sigma_2^\dagger a_k^\dagger
	\right]
	\Bigg\}|g,g,\mathrm{vac}\rangle,\label{wavefunction}
\end{align}
where $a_k^\dagger=\frac{1}{\sqrt{N}}\sum_n e^{ikn}a_n^\dagger$ 
and similarly for $b_k^\dagger$. Here, $c_e(t)$ denotes the probability amplitude for both emitters being excited, $c_D(n_1,n_2;t)$ describes the doublon sector, and $c_{ik}(t)  (i=1,2)$ is the probability amplitude 
for emitter $i$ and mode $k$ being simultaneously excited. We consider the initial state $|e,e,\mathrm{vac}\rangle$, where both emitters are excited and the lattice is in the vacuum state, i.e., $c_e(0)=1$.

After adiabatically eliminating the single-photon intermediate states (see Appendix~\ref{app:Derivation} for the detailed derivation), the system dynamics reduces to an effective description in which the two-emitter system can be described as an effective emitter with transition frequency $\omega_0=2\omega_e$, coupled to the dual SSH lattice that governs the doublon dynamics. The effective Hamiltonian can be written as
\begin{align}
	H_{\mathrm{eff}} &= H_{\mathrm{atom}}
	+ H_{\mathrm{bath}} + H_{\mathrm{int}},
	\\
	H_{\mathrm{atom}}
	&=\omega_0S^z,
	\\
	H_{\mathrm{bath}}
	&=
	\sum_{m=-N}^{N}
	V D_m^\dagger D_m
	\nonumber\\
	&-
	\sum_{n=-N/2}^{N/2}
	\left[
	J_1D_{2n}^\dagger D_{2n+1}+J_2D_{2n+1}^\dagger D_{2n+2}+\mathrm{H.c.}
	\right],\label{H_bath1}
	\\
	H_{\mathrm{int}}
	&=
	S^\dagger\left(g_0D_0+g_+D_1+g_-D_{-1}\right)+\mathrm{H.c.},
\end{align}
where $S^\dagger=\sigma_1^\dagger\sigma_2^\dagger$ and
$S^z=(|e,e\rangle\langle e,e|-|g,g\rangle\langle g,g|)/2$ are the
raising and Pauli-$z$ operators of the effective emitter, whose
transition frequency is $\omega_0=2\omega_e$, while $D_n^\dagger$
($D_n$) creates (annihilates) a doublon at site $n$ of the effective
dual lattice. The corresponding effective coupling amplitudes are
\begin{align}
	g_0&=\frac{g^2}{\sqrt{\omega_e^2-4J_{1}^{2}}}+\frac{g^2}{\sqrt{\omega_e^2-4J_{2}^{2}}},\label{g_01}
\\
g_+&=\frac{g^2\left( \omega_e -\sqrt{\omega_e ^2-4J_{1}^{2}} \right)}{2J_1\sqrt{\omega_e ^2-4J_{1}^{2}}},
\\
g_-&=\frac{g^2\left( \omega_e -\sqrt{\omega_e ^2-4J_{2}^{2}} \right)}{2J_2\sqrt{\omega_e ^2-4J_{2}^{2}}}.
\end{align}
We find that the coupling to the central site is dominant, with
$
g_0 \gg g_\pm 
$, which allows us to neglect the contributions from the neighboring sites $D_{\pm1}$. The interaction Hamiltonian then reduces to
\begin{equation}
	H_{\mathrm{int}} = g_0 S^\dagger D_0 + \mathrm{H.c.}, \label{Heff}
\end{equation}
so that the two-emitter system is mapped onto an effective emitter
coupled to the dual SSH lattice through the central doublon site
$D_0$.
\section{Vacancy-like Dressed States and Topological Bound States}
\label{sec:VDS}

We now characterize the bound state formed when the effective emitter frequency is tuned to the SSH edge-state energy, i.e., $\omega_0=E_{\mathrm{edge}}$. In this regime, the emitter frequency lies in the doublon band gap, so no propagating bulk doublon modes are available; nevertheless, a localized bound state can form around the emitter~\cite{Calajo2016,Wangxin2022,Guoshangjie2020,John1990}. We denote the excited and ground states of the effective emitter by $|e\rangle_{\mathrm{eff}}$ and $|g\rangle_{\mathrm{eff}}$, respectively, with
$|e\rangle_{\mathrm{eff}}\equiv|e,e\rangle$ and $|g\rangle_{\mathrm{eff}}\equiv|g,g\rangle$. The vacancy-like dressed state (VDS) theory~\cite{Leonforte2021} provides a natural framework for describing this bound state.

We write the VDS ansatz as
\begin{equation}
	|\Psi\rangle
	\propto
	\varepsilon |e\rangle_{\mathrm{eff}}|\mathrm{vac}\rangle
	+
	|g\rangle_{\mathrm{eff}}|\psi\rangle,\quad |\psi\rangle = \sum_n \psi_n |n\rangle,
\end{equation}
where \(|n\rangle = D_n^\dagger |\mathrm{vac}\rangle\) denotes a doublon localized at site \(n\) of the effective SSH lattice. The first term describes the effective emitter excitation, while the second term corresponds to the doublon wave function that dresses the emitter. Substituting this ansatz into the Schrödinger equation \(H_{\mathrm{eff}}|\Psi\rangle = \omega_0|\Psi\rangle\) and projecting onto the excited state of the effective emitter $\langle e,\mathrm{vac}|$, we obtain
\begin{equation}
	\omega_0 \varepsilon + g_0 \psi_{D_0}
	=
	\omega_0 \varepsilon, \label{vdscondition}
\end{equation}
which immediately yields \(\psi_{D_0} = 0\). This node condition is the defining feature of the VDS: the doublon wave function vanishes at the coupling site \(D_0\). Intuitively, the emitter and the doublon field interfere destructively at this point, which effectively removes the site \(D_0\) from the doublon lattice. The emitter thus behaves as a vacancy—a hard-wall boundary that the doublon wave function cannot penetrate.

Since $\psi_{D_0}=0$, the doublon dynamics can be mapped onto the
reduced lattice $B_v$, obtained by removing $D_0$ from the SSH chain. The doublon wave function therefore satisfies
\begin{equation}
	H_{B_v}|\psi\rangle=\omega_0|\psi\rangle,
	\label{vancy_Eq}
\end{equation}
where $H_{B_v}$ denotes the SSH Hamiltonian defined on the remaining
sites. The vacancy at $D_0$ effectively splits the SSH chain into two
segments with different boundary terminations. In the topological
regime $J_1<J_2$, only the segment with the appropriate termination
supports a vacancy-induced SSH edge mode. The resulting edge mode is
localized on the sublattice coupled to the effective emitter and thus
forms the VDS. In the thermodynamic limit, its doublon wave function~\cite{Sirker2014}
takes the form
\begin{align}
	\psi_{2n} &= 0, \\
	\psi_{2n+1}
	&=
	\frac{2\sqrt{\delta}}{1+\delta}
	\left(
	\frac{\delta-1}{\delta+1}
	\right)^n,
\end{align}
where \(n=0,1,\dots,(N-1)/2\) and
$
	\delta=J_2-J_1/J_1+J_2.
$
This wave function has three characteristic features. First, it decays exponentially away from the vacancy. Second, it has support only on one sublattice, reflecting the chiral symmetry of the SSH model. Third, it is localized exclusively on one side of the vacancy. The resulting single-sided localization, together with the sublattice polarization of the SSH edge mode, gives rise to the chiral character of the doublon bound state.

The direction of this single-sided localization is determined by the sublattice to which the emitter is coupled: coupling to an odd (or even) site yields a doublon wave function that decays to the right (or left). This sublattice-dependent chirality originates from the chiral symmetry of the SSH model and provides a means for engineering directional doublon-mediated interactions. This capability will be exploited in the four-body setup discussed in Sec.~\ref{Four-body}.

\begin{figure*}[ht]
	\centering \includegraphics[width=\textwidth]{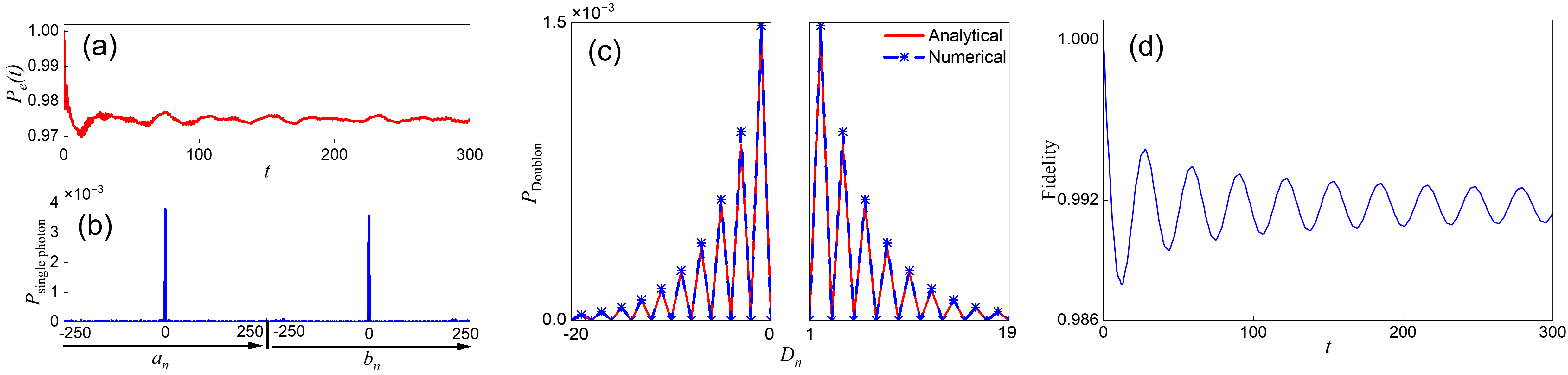}
	\caption{(a) Numerical results for the atomic excitation probability $P_e(t)$ as a function of time $t$. 
	(b) Spatial distribution of the single-photon component in the long-time limit. (c) Spatial distribution of the doublon component in the effective SSH chain. The left (right) panel corresponds to the emitter coupled to site $D_0$ ($D_1$).
	(d) Time-dependent fidelity between the numerically evolved state and the analytical VDS state.  
	The parameters are \(g = 0.3J_2\), \(\omega_e = 5J_2\) (so that \(\omega_0 = 10J_2\)), \(L = 1000\) sites, and all other parameters are the same as in Fig.~\ref{fig1}.
	}
	\label{fig2}
\end{figure*}

With the doublon eigenstate $|\psi\rangle$ determined, the corresponding
VDS is obtained by superposing the effective emitter excitation with the
doublon component. The normalized VDS can be written as
\begin{equation}
	|\Psi\rangle
	=
	\cos\theta\,|e\rangle_{\mathrm{eff}}|\mathrm{vac}\rangle
	+
	e^{i\phi}\sin\theta\,|g\rangle_{\mathrm{eff}}|\psi\rangle,
\end{equation}
with
\begin{equation}
	\theta=\arctan|\eta|,
	\quad
	\phi=\arg(\eta),
	\quad
	\eta
	=
	-\frac{g_0}
	{\langle D_0|H_{\mathrm{bath}}|\psi\rangle}.
\end{equation}
To verify the VDS picture, we numerically simulate the coupled emitter-dual-SSH system. Figure~\ref{fig2}(a) shows the effective emitter excitation probability $P_e(t)$, which approaches a finite value close to $|\cos\theta|^2$ at long times, providing a
signature of the vacancy-like dressed bound state. Figure~\ref{fig2}(b) shows the single-photon intermediate component, which remains localized around the emitter-coupling sites. Its short spatial extent indicates
that it mainly contributes to local dressing and has a negligible effect on the long-range interaction discussed below.

The spatial profile of the VDS is shown in Fig.~\ref{fig2}(c). The left and right panels show coupling at sites $D_0$ and $D_1$,
respectively. In both cases, the doublon wave function is exponentially localized, but its decay direction reverses with the coupling site:
it decays to the right for coupling at $D_0$ and to the left for coupling at $D_1$. This sublattice-dependent chirality reflects the sublattice-selective nature of the SSH edge states underlying the VDS.

To further validate the VDS description, we compute the fidelity
between the VDS ansatz and the state evolved under the full Hamiltonian.
We take the initial state to be
\begin{align}
	|\psi(0)\rangle
	&=
	\sqrt{P_e}\,
	|e\rangle_{\mathrm{eff}}|\mathrm{vac}\rangle
	\nonumber\\
	&+
	e^{i\phi}\sqrt{1-P_e}
	\sum_{n=0}^{(N-1)/2}
	c_{2n+1}\,
	|g\rangle_{\mathrm{eff}}|2n+1\rangle_D,
	\\
	c_{2n+1}
	&=
	\frac{2\sqrt{\delta}}{1+\delta}
	\left(
	\frac{\delta-1}{\delta+1}
	\right)^n,
\end{align}
where $P_e=|\cos\theta|^2$ is the effective emitter excitation
probability in the VDS. The system is then evolved under the full Hamiltonian $H$ defined in Eq.~(\ref{originalH}), and the time-dependent
fidelity is evaluated as
\begin{equation}
	\mathcal{F}(t)
	=
	\frac{
		|\langle\psi(0)|\psi(t)\rangle|^2
	}{
		\langle\psi(0)|\psi(0)\rangle
		\langle\psi(t)|\psi(t)\rangle
	},
\end{equation}
where $|\psi(t)\rangle$ denotes the numerically evolved state.

The time dependence of the fidelity is shown in Fig.~\ref{fig2}(d). We find that \(\mathcal{F}(t)\) remains close to unity throughout the evolution, with only weak oscillations around \(\mathcal{F} \simeq 0.992\). This high fidelity demonstrates that the VDS ansatz accurately captures the dominant bound-state component of the system. The small deviation from perfect fidelity can be attributed to residual single-photon intermediate-state contributions that are not fully captured by the effective VDS description.

\section{Four-body dipole-dipole interaction}
\label{Four-body}
\begin{figure}[ht]
	\centering \includegraphics[width=\columnwidth]{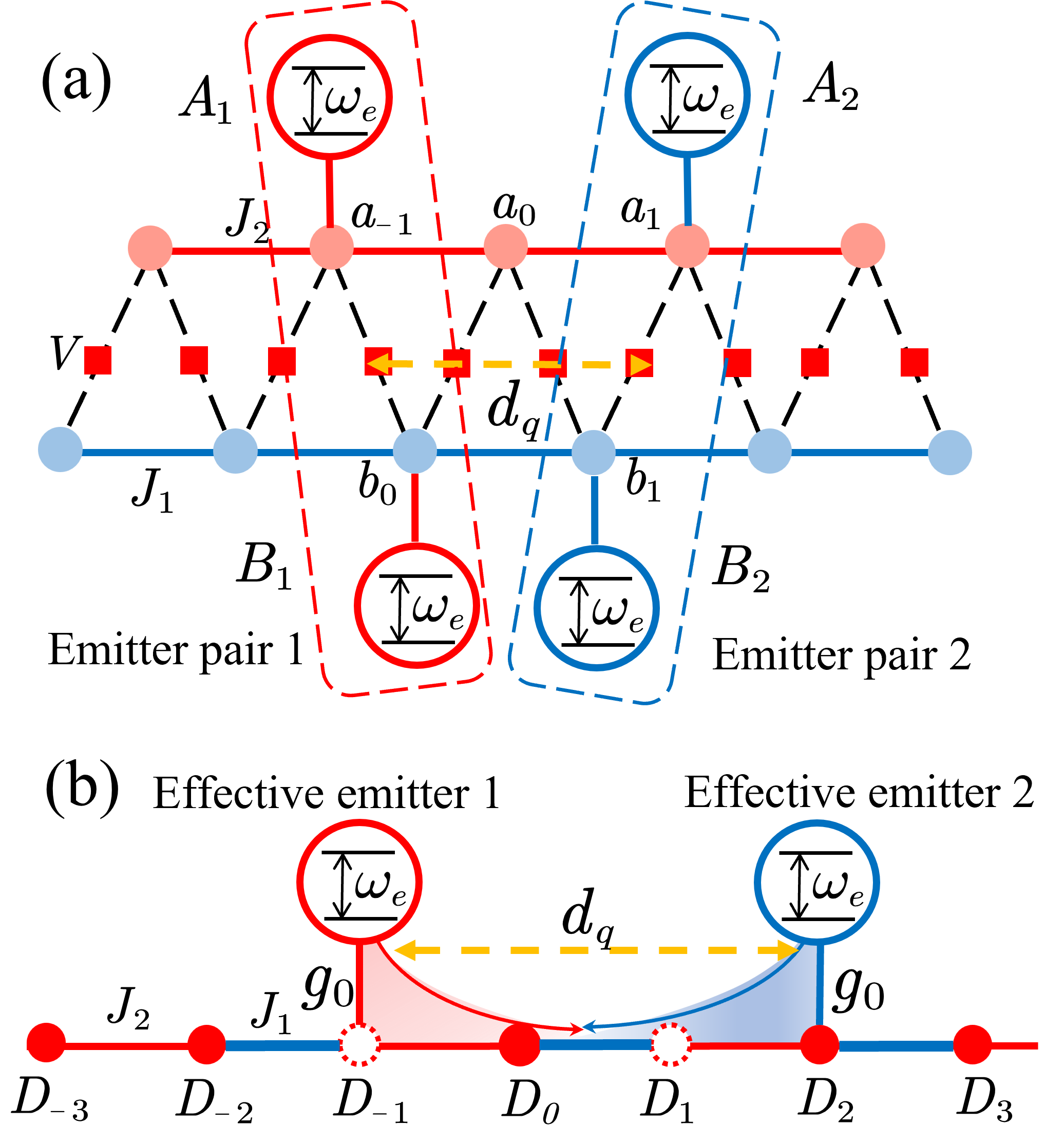}
	\caption{
		(a) Four-body setup consisting of two emitter pairs coupled to the triangular ladder waveguide. 
		(b) Effective model with two emitters coupled to the SSH chain at sites \(D_{-1}\) and \(D_2\), respectively; the separation between the two emitters is \(d_q = 3\).
	}
	\label{fig3}
\end{figure}

The chiral bound states identified in Sec.~\ref{sec:VDS} are exponentially localized and decay only to one side of the emitter. When the separation \(d_q\) between two such bound states is smaller than or comparable to their localization length, their tails overlap, giving rise to an effective dipole-dipole interaction~\cite{Wangxinlongrange,Bello2019,GonzlezTudela2015,Douglas2016,Santos2023}. We now explore this interaction by extending the setup to two emitter pairs, as shown in Fig.~\ref{fig3}(a). Each emitter pair consists of two physical emitters coupled to different legs of the ladder. The first pair \((A_1,B_1)\), with frequencies \((\omega_{A1},\omega_{B1})\), is coupled at sites \((a_{-1}, b_0)\), while the second pair \((A_2,B_2)\), with frequencies \((\omega_{A2},\omega_{B2})\), is coupled at sites \((a_1, b_1)\). According to the mapping established in Sec.~\ref{sec:Model}, each pair behaves as an effective emitter coupled to a single site of the dual SSH lattice, as illustrated in Fig.~\ref{fig3}(b).

The effective Hamiltonian for the two effective emitters can be written as
\begin{align}
	H_{\mathrm{eff},2} &= H_{\mathrm{atom,2}} + H_{\mathrm{bath}} + H_{\mathrm{int,2}}, \\
	H_{\mathrm{atom,2}} &= \omega_0 \left( S_1^z + S_2^z \right),
\end{align}
where \(H_{\mathrm{bath}}\) is the effective SSH Hamiltonian governing the doublon dynamics, given in Eq.~(\ref{H_bath1}), and the interaction between the effective emitters and the doublon lattice is
\begin{equation}
	H_{\mathrm{int,2}} = g_0 (S_1^\dagger D_x + S_2^\dagger D_{x+d_q}) + \mathrm{H.c.}
\end{equation}
Here, \(\omega_0 = \omega_{A1} + \omega_{B1} = \omega_{A2} + \omega_{B2}\) is the effective transition frequency associated with each emitter pair. To derive their effective interaction, we first diagonalize the bath Hamiltonian,
\begin{equation}
	H_{\mathrm{bath}}|\phi_m\rangle = \varepsilon_m |\phi_m\rangle,
\end{equation}
where \(\varepsilon_m\) and \(|\phi_m\rangle\) are the eigenvalues and eigenstates of the $H_{\mathrm{bath}}$. In this eigenbasis, the bath Hamiltonian becomes \(H_{\mathrm{bath}} = \sum_m \varepsilon_m c_m^\dagger c_m\), and the doublon annihilation operator at site \(j\) can be expanded as
\begin{equation}
	D_j = \sum_m \phi_m(j) c_m,
\end{equation}
with \(\phi_m(j) = \langle j|\phi_m\rangle\) denoting the wave function amplitude of the \(m\)th eigenmode at site \(j\).

In the interaction picture with respect to
$H_{\mathrm{bath}}+H_{\mathrm{atom,2}}$, the interaction Hamiltonian becomes~\cite{Soro2023,Huangyong2012,Shahmoon2013}
\begin{equation}
	H_I(t)
	=
	\sum_{i=1,2}\sum_m
	g_0 S_i^\dagger \phi_m(j_i)c_m
	e^{-i(\varepsilon_m-\omega_0)t}
	+\mathrm{H.c.},
\end{equation}
with $j_1=x$ and $j_2=x+d_q$. Since $\omega_0$ lies in the
doublon band gap, the bulk doublon modes are off-resonant and can be
adiabatically eliminated. Retaining the exchange term relevant to the
interaction between the two effective emitters gives
\begin{equation}
	H_{\mathrm{eff}}
	=
	\sum_m
	\frac{g_0^2}{\varepsilon_m-\omega_0}
	\left[
	\phi_m^*(x)\phi_m(x+d_q)
	S_1^\dagger S_2
	+\mathrm{H.c.}
	\right]
\end{equation}
Thus, the effective dipole-dipole interaction strength is
\begin{equation}
	J_{\mathrm{eff}}
	=
	g_0^2
	\sum_m
	\frac{\phi_m^*(x)\phi_m(x+d_q)}
	{\varepsilon_m-\omega_0},
	\label{Jeff}
\end{equation}
which is determined by the mode amplitudes at the two coupling sites
and their energy detunings from $\omega_0$. For $\omega_0$ in the
doublon band gap, the resulting interaction decays exponentially with
the separation $d_q$.

\subsection{``face-to-face'' configuration}
\begin{figure*}[ht]
	\centering \includegraphics[width=\textwidth]{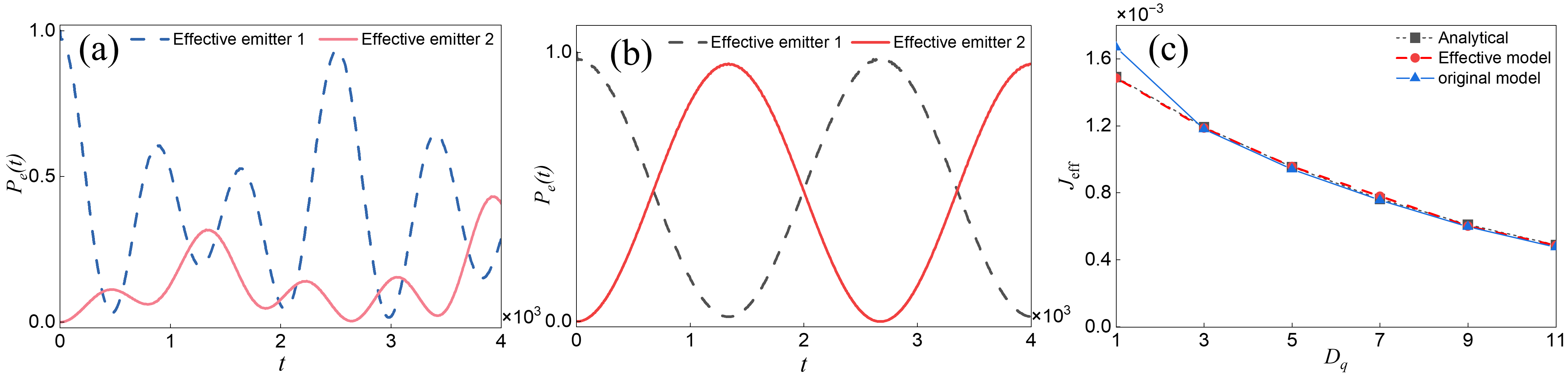}
	\caption{Population dynamics of the four-body system for separation $d_q=3$ in the ``face-to-face'' configuration. 
	(a) Dynamics for identical emitter frequencies, 
	$\omega_{A1}=\omega_{B1}=\omega_{A2}=\omega_{B2}=\omega_0/2$.  
	(b) Dynamics for finite detuning $\delta\omega=0.5J_2$.
	(c) Comparison between the numerically extracted oscillation frequencies and the analytical prediction of $J_{\mathrm{eff}}$ as a function of the separation $d_q$. 
	The parameter is $L=300$ sites and other parameters are consistent with Fig.~\ref{fig2}.
}
	\label{fig4}
\end{figure*}	
We first consider the face-to-face configuration, in which the two chiral bound states are oriented toward each other. Specifically, we consider two effective emitters coupled at sites \(D_{-1}\) and \(D_2\), with a separation \(d_q = 3\), as illustrated in Fig.~\ref{fig3}(b). In this arrangement, the bound state from the left emitter decays to the right, while that from the right emitter decays to the left. Their tails thus overlap, allowing coherent energy transfer between the two emitter pairs via the doublon lattice. 

Initially, the first emitter pair is prepared in the excited state while the second pair remains in the ground state. Figure~\ref{fig4}(a) shows the population dynamics for the case where all four emitters have the identical transition frequencies \(\omega_{A1} = \omega_{B1} = \omega_{A2} = \omega_{B2} = \omega_0/2\). Instead of the regular Rabi oscillations expected from the effective dipole-dipole interaction, the system exhibits irregular and aperiodic dynamics. This behavior arises from the coexistence of two distinct interaction pathways: the doublon-mediated exchange and single-photon exchange processes. The latter occur between emitters on the same leg of the ladder (\(A_1 \leftrightarrow A_2\) and \(B_1 \leftrightarrow B_2\)). The interference between these two channels suppresses the coherent exchange dynamics.

To isolate the doublon-mediated dipole-dipole interaction, we introduce a frequency detuning that suppresses the single-photon exchange channels while preserving the resonance condition for doublon exchange between the two emitter pairs. The emitter frequencies are chosen as
\begin{align}
	\omega_{A1} &= \omega_{B1} = \frac{\omega_0}{2}, \\
	\omega_{A2} &= \omega_{A1} + \delta\omega, \\
	\omega_{B2} &= \omega_{B1} - \delta\omega,
\end{align}
so that both emitter pairs retain the same total transition energy,
\begin{equation}
	\omega_0
	=
	\omega_{A2}+\omega_{B2}
	=
	\omega_{A1}+\omega_{B1}.
\end{equation}
Thus, the doublon exchange remains resonant, whereas the single-photon exchange channels are detuned by $\pm\delta\omega$, thereby suppressing their contribution to the dynamics.

In addition to the resonant doublon-exchange channel, virtual single-photon processes also contribute. Although off-resonant, these processes induce a frequency renormalization of the emitters, corresponding to a Stark shift~\cite{Scully1997,Valente2017}. Since the effective coupling $J_{\mathrm{eff}}$ in Eq.~(\ref{Jeff}) is derived using the unshifted emitter frequencies, this correction must be taken into account in the numerical simulations. The Stark shift of each emitter is given by
\begin{align}
	\Delta\omega_{A(B),i}
	&=
	-\frac{1}{N}\sum_k
	\frac{g^2}{\delta_{A(B),i}(k)}, \\
	\delta_{A(B),i}(k)
	&=
	\omega_{A(B),i}-\epsilon_i(k),
\end{align}
where $\epsilon_i(k)$ denotes the single-photon dispersion of the corresponding ladder leg. We then adjust the bare emitter frequencies used in the numerical simulations to compensate for this shift, such that the resulting dressed frequencies remain at their target values.

Figure~\ref{fig4}(b) shows the population dynamics for $\delta\omega=0.5J_2$ after accounting for the Stark-shift correction. In contrast to Fig.~\ref{fig4}(a), the dynamics now exhibit coherent Rabi oscillations between the two emitter pairs, with an oscillation frequency consistent with the analytical prediction of the effective Hamiltonian. The excitation is coherently exchanged through the doublon-mediated interaction, confirming that the single-photon exchange channels have been effectively suppressed. As shown in Fig.~\ref{fig4}(c), we further extract the oscillation frequencies from numerical simulations of both the original triangular-ladder model and the effective SSH doublon model for different separations $d_q$. The numerical results agree well with the analytical prediction of $J_{\mathrm{eff}}$ obtained from the effective Hamiltonian, confirming the validity of the effective description.

\subsection{``back-to-back'' configuration}
\begin{figure}[ht]
	\centering \includegraphics[width=\columnwidth]{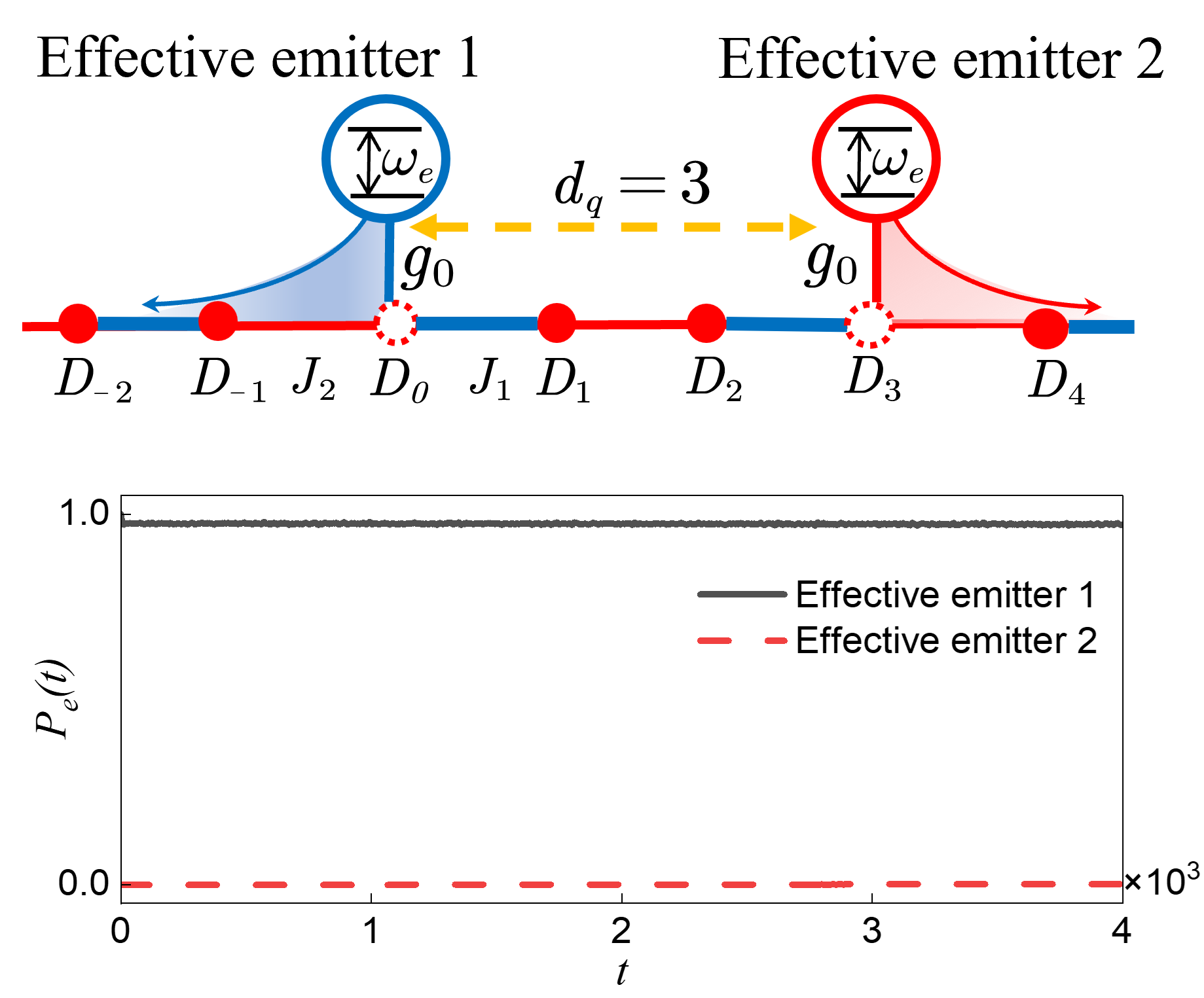}
	\caption{Population dynamics of the four-body system for the ``back-to-back'' configuration with a separation \(d_q = 3\). The detuning is set to \(\delta\omega = 0.5J_2\), and other parameters are consistent with Fig.~\ref{fig4}.
	}
	\label{fig5}
\end{figure}	
We now consider the back-to-back configuration, in which the two chiral bound states are oriented away from each other. Specifically, we consider two effective emitters coupled at sites \(D_0\) and \(D_3\), with a separation \(d_q = 3\). The emitter at the even site \(D_0\) gives rise to a bound state that decays to the left, while the emitter at the odd site \(D_3\) gives rise to one that decays to the right. The two bound states therefore reside on opposite sides of the lattice, with vanishing amplitude in the region between them. Their spatial overlap is thus suppressed.

After introducing the detuning $\delta\omega = 0.5J_2$ to suppress single-photon exchange processes, the population dynamics are shown in
Fig.~\ref{fig5}. The populations of both emitter pairs remain unchanged throughout the evolution, indicating the absence of appreciable coherent excitation transfer. This result contrasts with the face-to-face case in Fig.~\ref{fig4}(b), where the same detuning gives rise to coherent Rabi oscillations. The comparison thus confirms that the interaction depends not only on the emitter separation but also on the relative chirality of the bound states.

\subsection{Case of identical chirality: with even separation}
\begin{figure}[ht]
	\centering \includegraphics[width=\columnwidth]{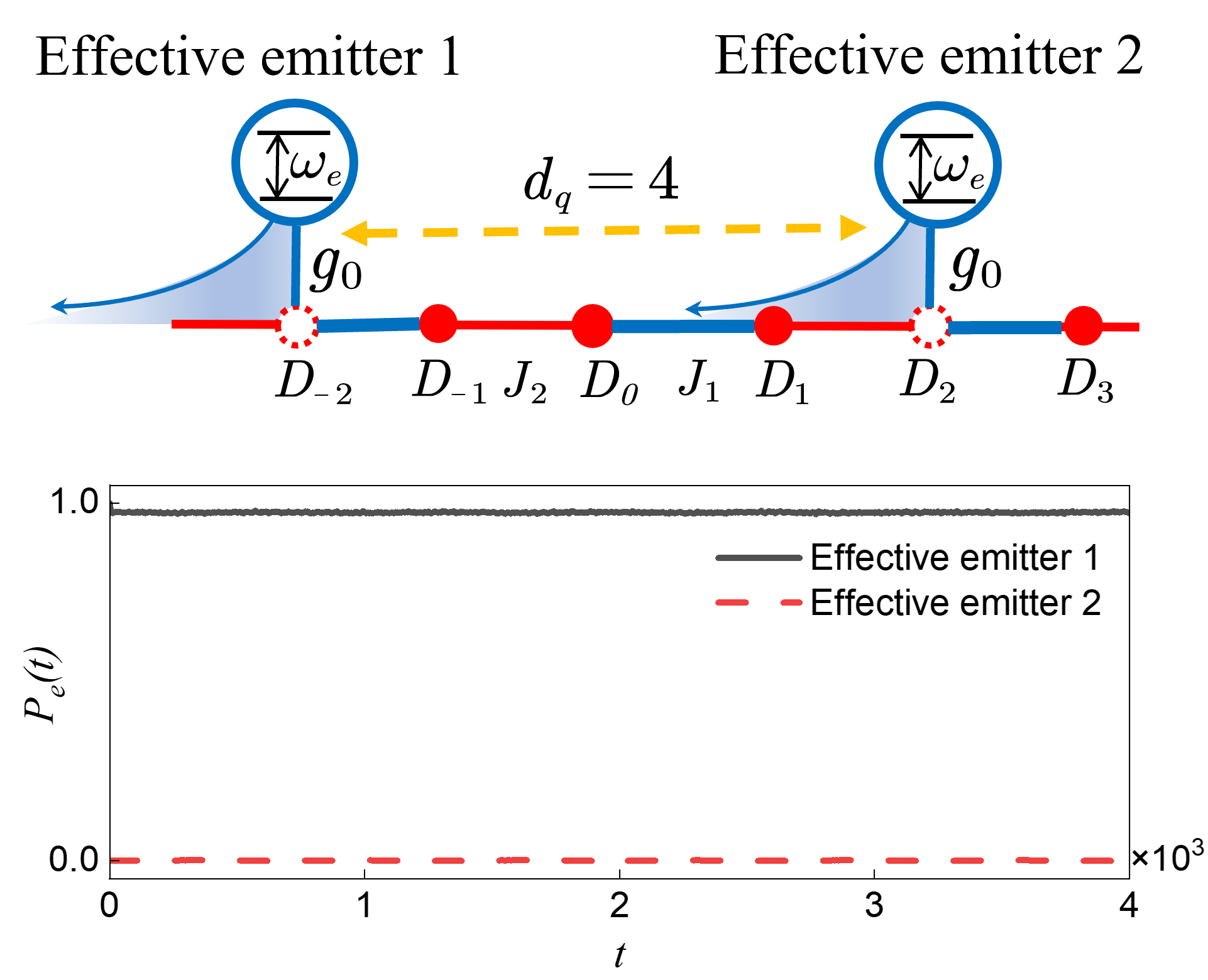}
	\caption{Population dynamics of the four-body system for the ``same-chirality'' configuration with a separation \(d_q = 4\). The detuning is set to \(\delta\omega = 0.5J_2\), and other parameters are consistent with Fig.~\ref{fig4}.
	}
	\label{fig6}
\end{figure}	
We now consider the case where the two effective emitters are coupled to sites belonging to the same sublattice, so that their chiral bound states decay in the same direction. As an example, we choose the coupling sites to be $D_{-2}$ and $D_2$, with a separation $d_q=4$. As discussed in Sec.~\ref{sec:VDS}, the chiral doublon bound state is sublattice selective and has support only on one sublattice. In this configuration, both effective emitters are coupled to even sites. Consequently, the bound state associated with the first emitter has
vanishing amplitude at the second coupling site, and vice versa. This sublattice selectivity eliminates the overlap between the two bound states, leading to $J_{\mathrm{eff}}=0$.

With the detuning $\delta\omega = 0.5J_2$ introduced to suppress single-photon exchange processes, the population dynamics for this configuration are shown in Fig.~\ref{fig6}. The populations of both emitter pairs remain essentially unchanged throughout the evolution, indicating the absence of appreciable coherent excitation transfer. Together with the face-to-face and back-to-back results, these observations demonstrate that the effective interaction depends not only on the emitter separation but also on the sublattice structure and relative chirality of the bound states.
\section{Conclusion}
In this work, we have shown that a triangular ladder waveguide with strong nearest-neighbor interactions can support a chiral doublon bound state. The doublon dynamics can be mapped onto an effective SSH chain. Using the vacancy-like dressed state framework, we identify a bound state with a node at the coupling site. This state inherits the sublattice selectivity and exponential localization of the SSH edge states.

Extending this picture to two emitter pairs, we find that the effective four-body dipole-dipole interaction depends on the relative chirality of the bound states and the parity of their separation. When the two bound states face each other, their spatial overlap gives rise to coherent Rabi oscillations between the two emitter pairs. In the back-to-back configuration, their spatial overlap vanishes, suppressing the effective interaction. When the two emitter pairs are coupled to the same sublattice with an even separation, the interaction also vanishes because the bound states have support only on a single sublattice. The numerical results agree with the analytical predictions of the effective theory.

The required strong nearest-neighbor interactions can be realized using tunable nonlinear couplers in circuit QED (see Appendix~\ref{app:circuit} for a possible circuit implementation), providing a feasible route toward experimental realization of the proposed scheme. These results demonstrate that chirality can be used to control doublon-mediated interactions between correlated photon pairs. This mechanism could be exploited to engineer tailored many-body quantum systems, with potential applications in quantum simulation and the design of long-range quantum interactions.

\section{Acknowledgments}
The quantum dynamical simulations are based on open source code 
QuTiP. 
X.W. is supported by Natural Science Basic Research Program of Shaanxi Province (Grant No. 2026JC-YXQN-027), Shaanxi Fundamental Science Research Project for Mathematics and Physics
(Grant No. 25JSQ026), and the National Natural Science Foundation of China (NSFC) (Grant No. 12174303).

\appendix
\section{Derivation of the effective doublon Hamiltonian}
\label{app:Derivation}

We provide a detailed derivation of the effective Hamiltonian introduced in Sec.~\ref{sec:Model}. Starting from the wave-function ansatz in Eq.~(\ref{wavefunction}) and substituting it into the Schrödinger equation
$i\frac{d}{dt}|\psi(t)\rangle=H|\psi(t)\rangle$, we obtain
\begin{align}
	i\dot{c}_e
	&=
	2\omega_e c_e
	+
	\frac{g}{\sqrt{N_c}}
	\sum_k
	\left(c_{1k}+c_{2k}\right),
	\label{app_dotce}
	\\
	i\dot{c}_{1k}
	&=
	\delta_{k1}c_{1k}
	+
	\frac{g}{\sqrt{N_c}}
	\left(
	c_e+c_{D_0}+e^{-ik}c_{D_1}
	\right),
	\\
	i\dot{c}_{2k}
	&=
	\delta_{k2}c_{2k}
	+
	\frac{g}{\sqrt{N_c}}
	\left(
	c_e+c_{D_0}+e^{ik}c_{D_{-1}}
	\right),
	\\
	i\dot{c}_{D_0}
	&=
	Vc_{D_0}
	+
	\frac{g}{\sqrt{N_c}}
	\sum_k
	\left(c_{1k}+c_{2k}\right)
	-J_1c_{D_1}
	-J_2c_{D_{-1}},
	\label{app_dotcD0}
	\\
	i\dot{c}_{D_1}
	&=
	Vc_{D_1}
	+
	\frac{g}{\sqrt{N_c}}
	\sum_k e^{ik}c_{1k}
	-J_1c_{D_0}
	-J_2c_{D_2},
	\label{app_dotcD1}
	\\
	i\dot{c}_{D_{-1}}
	&=
	Vc_{D_{-1}}
	+
	\frac{g}{\sqrt{N_c}}
	\sum_k e^{-ik}c_{2k}
	-J_1c_{D_{-2}}
	-J_2c_{D_0},
	\label{app_dotcDm1}
	\\
	i\dot{c}_{D_2}
	&=
	Vc_{D_2}
	-J_1c_{D_3}
	-J_2c_{D_1},
	\\
	i\dot{c}_{D_{-2}}
	&=
	Vc_{D_{-2}}
	-J_1c_{D_{-1}}
	-J_2c_{D_{-3}},
\end{align}
where where $\delta_{ki} = \omega_e - 2J_i \cos k$ ($i=1,2$) denotes the detuning between the emitter transition frequency and the corresponding single-photon dispersion.

The intermediate single-photon states are
\begin{equation}
	|1,k\rangle
	=
	\sigma_1^\dagger b_k^\dagger
	|g,g,\mathrm{vac}\rangle,
	\quad
	|2,k\rangle
	=
	\sigma_2^\dagger a_k^\dagger
	|g,g,\mathrm{vac}\rangle.
\end{equation}
In the weak-coupling regime, these states are far off resonance $|\delta_{ki}|\gg g/\sqrt{N_c}$. They can therefore be adiabatically eliminated by setting $\dot{c}_{1k}=\dot{c}_{2k}=0$, which gives
\begin{align}
	c_{1k}
	&=
	-\frac{g}{\sqrt{N_c}\,\delta_{k1}}
	\left(
	c_e+c_{D_0}+e^{-ik}c_{D_1}
	\right),
	\label{app_c1k}
	\\
	c_{2k}
	&=
	-\frac{g}{\sqrt{N_c}\,\delta_{k2}}
	\left(
	c_e+c_{D_0}+e^{ik}c_{D_{-1}}
	\right).
	\label{app_c2k}
\end{align}

Substituting Eqs.~(\ref{app_c1k}) and (\ref{app_c2k}) into Eqs.~(\ref{app_dotcD0}-\ref{app_dotcDm1}), and replacing $\sum_k$ with $\frac{N_c}{2\pi}\int{dk}
$, we obtain 
\begin{align}
	i\dot{c}_e&=2\omega _ec_e-g_0c_{D_0}-g_{+}c_{D_1}-g_{-}c_{D_{-1}},
	\\
	i\dot{c}_{D_0}&=Vc_{D_0}-g_0c_e-J_1c_{D_1}-J_2c_{D_{-1}},
	\\
	i\dot{c}_{D_1}&=Vc_{D_1}-g_+c_e-J_1c_{D_0}-J_2c_{D_2},
	\\
	i\dot{c}_{D_{-1}}&=Vc_{D_{-1}}-g_-c_e-J_1c_{D_{-2}}-J_2c_{D_0},
	\\
	i\dot{c}_{D_2}&=Vc_{D_2}-J_1c_{D_3}-J_2c_{D_1},
	\\
	i\dot{c}_{D_{-2}}&=Vc_{D_{-2}}-J_1c_{D_{-1}}-J_2c_{D_{-3}},
\end{align}
where the coupling strengths are given by
\begin{align}
	g_0 &= \frac{g^2}{N_c} \sum_k \left( \frac{1}{\delta_{k1}} + \frac{1}{\delta_{k2}} \right)
	= \frac{g^2}{\sqrt{\omega_e^2 - 4J_1^2}} + \frac{g^2}{\sqrt{\omega_e^2 - 4J_2^2}}, \label{g_0} \\
	g_+ &= \frac{g^2}{N_c} \sum_k \frac{e^{ik}}{\delta_{k1}}
	= \frac{g^2 \left( \omega_e - \sqrt{\omega_e^2 - 4J_1^2} \right)}{2J_1 \sqrt{\omega_e^2 - 4J_1^2}}, \\
	g_- &= \frac{g^2}{N_c} \sum_k \frac{e^{-ik}}{\delta_{k2}}
	= \frac{g^2 \left( \omega_e - \sqrt{\omega_e^2 - 4J_2^2} \right)}{2J_2 \sqrt{\omega_e^2 - 4J_2^2}}.
\end{align}

The above procedure yields an effective description in which the two local emitters form a single effective emitter with transition frequency $\omega_0=2\omega_e$, coupled to the dual SSH lattice governing the doublon dynamics. The effective Hamiltonian can be written as
\begin{align}
	H_{\mathrm{eff}} &= H_{\mathrm{atom}}
	+ H_{\mathrm{bath}} + H_{\mathrm{int}},
	\\
	H_{\mathrm{atom}}
	&=\omega_0S^z,
	\\
	H_{\mathrm{bath}}
	&=
	\sum_{m=-N}^{N}
	V D_m^\dagger D_m
	\nonumber\\
	&-
	\sum_{n=-N/2}^{N/2}
	\left[
	J_1D_{2n}^\dagger D_{2n+1}+J_2D_{2n+1}^\dagger D_{2n+2}+\mathrm{H.c.}
	\right],\label{H_bath}
	\\
	H_{\mathrm{int}}
	&=
	S^\dagger\left(g_0D_0+g_+D_1+g_-D_{-1}\right)+\mathrm{H.c.}
\end{align}
Since $g_0\gg g_\pm$, the couplings to the neighboring sites $D_{\pm1}$ can be neglected, yielding Eq.~(\ref{Heff}) of the main text.

\section{Circuit realization of the cross-Kerr interaction}
\label{app:circuit}
\begin{figure}[ht]
	\centering \includegraphics[width=\columnwidth]{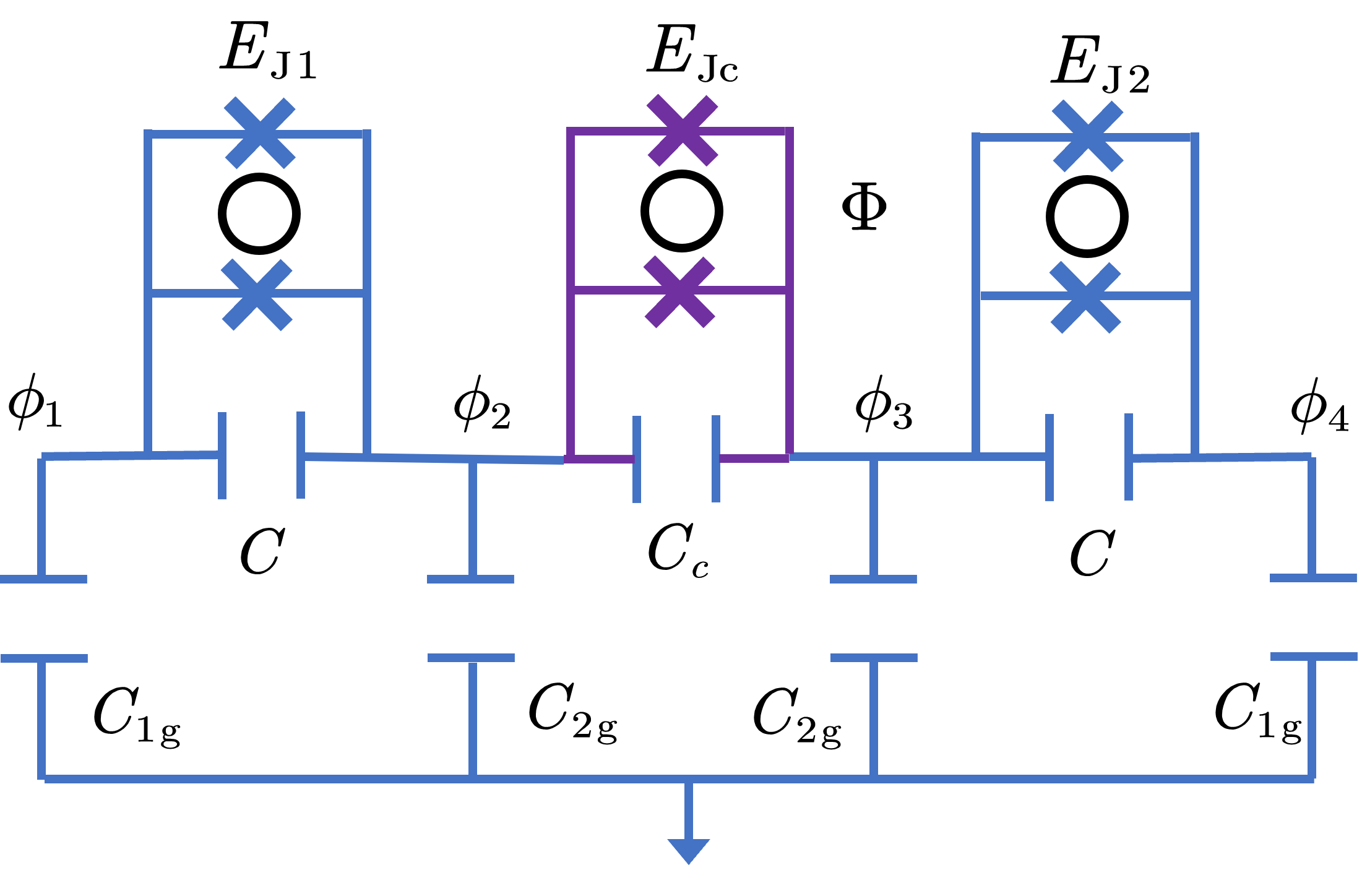}
	\caption{Experimental setup for the cross‑Kerr interaction in the tunable coupler circuit. Two transmon qubits are coupled via a capacitor $C_c$ and a flux-tunable SQUID. }
	\label{fig7}
\end{figure}	
The cross-Kerr interaction considered in the main text can be realized using the nonlinear coupler introduced in Ref.~\cite{Kounalakis2018,Stolyarov2025,HuYong2011,JinJiasen2013,Jinjiasen2014}. The circuit consists of two transmons~\cite{Schreier2008,KochJens2007}, with Josephson energies \(E_J^{(1)}\) and \(E_J^{(2)}\), respectively, and each with capacitance \(C\), coupled via a capacitor \(C_c\) in parallel with a flux-tunable superconducting quantum
interference device (SQUID)~\cite{vanOtterlo1995,Tinkham1996}, as illustrated in Fig.~\ref{fig7}. The charging energy is defined as
\begin{equation}
	E_C = \frac{e^2}{2C}.
	\label{eq:app_EC}
\end{equation}
The capacitor provides a fixed linear capacitive coupling, while the SQUID acts as a flux-tunable nonlinear inductive element. For a symmetric SQUID, the effective Josephson energy is
\begin{equation}
	E_J^c(\Phi)
	=
	E_J^{c,\mathrm{max}}
	\left|
	\cos\left(\pi \frac{\Phi}{\Phi_0}\right)
	\right|,
	\quad
	\Phi_0 = \frac{h}{2e},
	\label{eq:app_EJc}
\end{equation}
where \(\Phi\) is the external magnetic flux threading the SQUID loop.

Denoting the node flux variables by \(\phi_i\) (\(i = 1,2,3,4\)), we introduce the dimensionless phase differences across the two transmons,
\begin{equation}
	\psi_A
	=
	\frac{2\pi}{\Phi_0}(\phi_1 - \phi_2),
	\quad
	\psi_B
	=
	\frac{2\pi}{\Phi_0}(\phi_4 - \phi_3),
	\label{eq:app_transmon_phases}
\end{equation}
together with the coupler mode coordinate
\begin{equation}
	\psi_S
	=
	\frac{\pi}{\Phi_0}
	(\phi_1 + \phi_2 - \phi_3 - \phi_4).
	\label{eq:app_coupler_mode}
\end{equation}
The phase difference across the coupling SQUID is then
\begin{equation}
	\psi_C
	=
	\frac{\psi_A - \psi_B}{2} - \psi_S.
	\label{eq:app_phase_relation}
\end{equation}
The inductive energy of the nonlinear circuit is therefore
\begin{align}
	\mathcal{E}_{\mathrm{ind}}
	=&
	-E_J^{(1)}\cos\psi_A
	-E_J^{(2)}\cos\psi_B
	\nonumber\\
	&
	-E_J^c(\Phi)
	\cos\psi_C.
	\label{appendix_inductive}
\end{align}
The first two terms describe the individual transmons, while the third term originates from the tunable coupler and generates interactions between them.

In the weakly coupled regime, the two local transmon modes are quantized as
\begin{equation}
	\psi_A = \left( \frac{2E_C}{E_J^{(1)}} \right)^{1/4} \left( \hat a + \hat a^\dagger \right), \quad
	\psi_B = \left( \frac{2E_C}{E_J^{(2)}} \right)^{1/4} \left( \hat b + \hat b^\dagger \right).
	\label{eq:app_quantization}
\end{equation}
Expanding the coupler potential
\(-E_J^c(\Phi)\cos\psi_C\) around \(\psi_C=0\), and omitting the
constant term, gives
\begin{equation}
	-E_J^c(\Phi)\cos\psi_C
	\simeq
	\frac{E_J^c(\Phi)}{2}\psi_C^2
	-
	\frac{E_J^c(\Phi)}{24}\psi_C^4
	+\cdots .
	\label{eq:app_Josephson_expand}
\end{equation}
When the two transmons are resonant (\(E_J^{(1)} = E_J^{(2)} \equiv E_J\)), the quadratic contribution, under the rotating-wave approximation, yields the linear hopping between the two modes,
\begin{equation}
	H_{\mathrm{hop}}
	=
	J
	\left(
	\hat a^\dagger \hat b +
	\hat b^\dagger \hat a
	\right).
	\label{eq:app_hop}
\end{equation}
The hopping amplitude is given by
\begin{equation}
	J = J_C - J_L(\Phi),
	\label{eq:app_J}
\end{equation}
where \(J_C\) and \(J_L(\Phi)\) are the capacitive and inductive contributions, respectively. At the filtering flux \(\Phi_{\mathrm{filter}}\), these two contributions satisfy~\cite{Geller2015}
\begin{equation}
	J_L(\Phi_{\mathrm{filter}}) = J_C,
	\quad
	J(\Phi_{\mathrm{filter}}) \simeq 0.
	\label{eq:app_Jzero}
\end{equation}

The quartic contribution generates nonlinear interactions between the two transmon modes. Its leading number-conserving cross term gives rise to the effective cross-Kerr interaction
\begin{equation}
	H_{\mathrm{cross-Kerr}}
	=
	V \,
	\hat a^\dagger \hat a \,
	\hat b^\dagger \hat b,
	\label{eq:app_crossKerr}
\end{equation}
with the leading-order coefficient given by~\cite{Kounalakis2018}
\begin{equation}
	V
	\simeq
	-\frac{E_J^c(\Phi) E_C}{8 E_J}.
	\label{eq:app_V}
\end{equation}
Thus, the cross-Kerr interaction is tunable via the external flux. Importantly, \(E_J^c(\Phi_{\mathrm{filter}})\) remains finite at the filtering point, so that the linear hopping can be suppressed while the nonlinear interaction persists. The circuit can therefore access the interaction-dominated regime
\begin{equation}
	|J| \ll |V|,
	\label{eq:app_interaction_dominated}
\end{equation}
which is precisely the regime required for the doublon dynamics studied in the main text.

By periodically repeating this nonlinear coupler architecture, the same circuit design can be scaled up to the triangular-ladder geometry considered in the main text. In this extended lattice, the cross-Kerr couplings realize the nearest-neighbor interaction \(V\), while the linear couplings along the two legs yield the hopping amplitudes \(J_1\) and \(J_2\).

%

\end{document}